\documentclass[11pt,a4paper]{article}
\usepackage{amsthm,amssymb,epsfig,latexsym,amsmath}
\newcommand{\be}[1]{\begin{equation}\label{#1}}
\newcommand{\ba}[1]{\begin{eqnarray}\label{#1}}
\newcommand{\ee}{\end{equation}}
\newcommand{\ea}{\end{eqnarray}}

\def\det{\operatorname{det}}

\newcommand{\bra}[1]{\langle\,#1\,|}
\newcommand{\ket}[1]{|\,#1\,\rangle}

\newtheorem{theorem}{Theorem}[section]

\def\beqa{\begin{eqnarray}}
\def\eeqa{\end{eqnarray}}
\def\ba{\begin{array}}
\def\ea{\end{array}}
\def\r{\rangle}
\def\l{\langle}

\def\b{\beta}

\def\e{\epsilon}

\def\la{\lambda}
\def\s{\sigma}
\def\sul{\sum\limits}
\def\pl{\prod\limits}
\def\lt({\left(}
\def\rt){\right)}
\def\pd #1{\frac{\partial}{\partial #1}}

\newtheorem{lemma}{Lemma}[section]
\def\qed{\hfill\nobreak\hbox{$\square$}\par\medbreak}
\makeatletter
\@addtoreset{equation}{section}
\makeatother

\begin{document}
\begin{flushright}
LPENSL-TH-03/04\\
\end{flushright}
\par \vskip .1in \noindent

\vspace{24pt}

\begin{center}
\begin{LARGE}
{\bf On the spin-spin correlation functions
 of the $XXZ$ spin-$\textstyle{\frac{1}{2}}$ infinite
chain} \end{LARGE}

\vspace{50pt}

\begin{large}

{\bf N.~Kitanine}\footnote[1]{LPTM, UMR 8089 du CNRS, Universit\'e de Cergy-Pontoise,
France, 
kitanine@ptm.u-cergy.fr \par
\hspace{2mm} On leave of absence from Steklov Institute at
St. Petersburg, Russia},~~
{\bf J.~M.~Maillet}\footnote[2]{ Laboratoire de Physique, UMR 5672 du CNRS,
ENS Lyon,  France,
 maillet@ens-lyon.fr},~~
{\bf N.~A.~Slavnov}\footnote[3]{ Steklov Mathematical Institute,
Moscow, Russia, nslavnov@mi.ras.ru},~~
{\bf V.~Terras}\footnote[4]{LPMT, UMR 5825 du CNRS,
Montpellier, France, terras@lpm.univ-montp2.fr}.
\end{large}

\vspace{70pt}

\centerline{\bf Abstract} \vspace{1cm}
\parbox{12cm}{\small We obtain  a new
  multiple integral representation  for
 the spin-spin correlation functions of the XXZ spin-$\frac 12$ infinite chain.
 We show that this representation
is closely related with the partition function of the six-vertex model with 
domain wall boundary conditions. }
\end{center}

\vspace{30pt}

\newpage
\section{Introduction}

The calculation of the correlation functions of the spin chains 
and, in particular,  their asymptotic analysis are very important problems in the field
of  quantum integrable
models. In this article we consider one of the most representative examples 
of the lattice integrable models: the spin-$\frac 12$ Heisenberg chain. This model 
describes spin  $\frac 12$ particles situated  in the sites of a one-dimensional
 lattice, interacting with their nearest neighbours,  
\be{IHamXXZ}
H=\sum_{m=1}^{M}\left(
\sigma^x_{m}\sigma^x_{m+1}+\sigma^y_{m}\sigma^y_{m+1}
+\Delta(\sigma^z_{m}\sigma^z_{m+1}-1)\right).
\ee
Here $\Delta$ is the anisotropy parameter and
 $\sigma^{x,y,z}_{m}$ are Pauli matrices,  associated with each site of
the chain. This Hamiltonian acts in a tensor product of $M$ two-dimensional local quantum
spaces $\mathcal{H}_m$. 
\footnote[1]{All the 
results of the present paper (with slight modifications) can also be applied
 to the XXZ model with external 
magnetic field. However for simplicity reason we  consider here only the case of zero magnetic field.}
We imposed here the periodic boundary conditions
\[\sigma^a_{M+1}=\sigma^a_1.\] For simplicity reason we consider here only the case
when  the number of sites $M$ is even. 

Our main goal is to obtain some explicit expression for the two-point spin-spin correlation
functions for the infinite chain at zero temperature.
 They can be defined as ground state mean values of 
products of local operators:
\begin{align}
\label{Defzz}
g_{z z}(m)=&\bra{\psi_g}\sigma^z_{m+1}\sigma^z_1\ket{\psi_g},\\
g_{+ -}(m)=&\bra{\psi_g}\sigma^+_{m+1}\sigma^-_1\ket{\psi_g},\label{Def+-}
\end{align}
where $\ket{\psi_g}$ is the ground state for the Hamiltonian (\ref{IHamXXZ}).

To solve this  model different  techniques can be used. Originally  the energy level
were calculated  in 1958 
by means of the  coordinate Bethe Ansatz \cite{Bet31,Orb58,Wal59,YanY66a}. 
Later (in 1979),  an algebraic version of the Bethe Ansatz
 ({\it or quantum  inverse scattering method})
was introduced by Faddeev, Sklyanin and Takhtajan
\cite{FadST79}. Both methods permit to calculate the energy level
and to obtain representations for the eigenstates and, in particular,
 for the ground state. However, for many years, computation
of the correlation functions  has been possible  only for the free fermion
point $\Delta=0$ \cite{LieSM61,Wu66,Mcc68,BarMTW76,ColIKT92}. 

First explicit results for more general situations, namely for the massive antiferromagnetic 
regime  ($\Delta>1$), were obtained in 1992 by means of a completely different
 approach ($q$-vertex operator method) by the Kyoto group  \cite{JimMMN92,JimM95L}. In 1996 a similar
conjecture was written for the critical regime of the XXZ chain ($-1<\Delta\le 1$)
\cite{JimM96}.  The correlation functions were represented as multiple integrals with number of
integrals equal to the distance. These 
results were confirmed (and generalised for the XXZ chain in a constant external magnetic 
field) in 1999 \cite{KitMT99,KitMT00} by means of  the algebraic Bethe Ansatz and resolution
of the {\it quantum inverse problem} \cite{KitMT99,MaiT99}. This new approach permitted to
understand better the results for the correlation functions and  to obtain an 
asymptotic formula for a very particular correlation function (the so-called 
{\it emptiness formation probability}, which is the probability to find a ferromagnetic
string of length $m$ in the ground state) \cite{KitMST02efp,KorLSN03}, and even an explicit result
for this quantity for the point $\Delta=\frac 12$ \cite{KitMST02efpP3}.

It is necessary to underline that the correlation functions calculated in 
\cite{JimMMN92,KitMT00} are not the two-point functions defined by (\ref{Defzz}) and 
(\ref{Def+-})
but  they are the so-called ``elementary blocks'' (i.e. the mean values of products of local elementary 
$2\times 2$ matrices with only one non-zero entry in consecutive sites). The two-point 
functions can be expressed as  sums of these elementary blocks but the number of terms 
in this sum grows exponentially with respect to the distance
($2^{m+1}$), which is very inconvenient for the computation
of the long distance asymptotics. The main goal of the present paper is to reduce our expression for the 
two-point functions
to a sum  of  $m+1$ terms and thus to obtain a compact and manageable
representation for it. One way to solve this problem was proposed in  \cite{KitMST02a}, and 
here we present an alternative way to simplify the expressions for the spin-spin correlation functions.
The advantage of this alternative approach is a simpler final form of the
 terms (each of them contains the
 same number
of integrals and there is no need to introduce any additional integrations). It is also important to
mention that each term in the final sum  has a form very similar to the symmetric representation for 
the emptiness formation probability, obtained in \cite{KitMST02a}, and 
thus there is a possibility to use the saddle point method similarly
to  \cite{KitMST02efp} for its rough asymptotic analysis.

The paper is organised as follows. In the next section we give a short reminder of the 
Algebraic Bethe Ansatz solution of the XXZ chain (following \cite{FadST79}) and 
an overview of our way to calculate the elementary blocks 
following our paper
\cite{KitMT00}. In  Section 3 we introduce the generating function 
of the correlation functions of the third components of spin
and we show how to 
express this quantity in terms of the elementary
blocks. We also explain how this function is related with the two-point
function (\ref{Defzz}). In  Section 4 we prove our main result for the generating 
function and show how a similar technique can be applied directly to the correlation functions
(\ref{Def+-}) and (\ref{Defzz}). A further re-summation is performed in  
Section 5, which permits us to reduce the result to only one term. 
In the last section we show how this method can be applied in 
the free fermion point to obtain the known results for the two-point functions and their asymptotics.     
\section{XXZ chain: algebraic Bethe Ansatz and elementary blocks}

 First we have to determine the ground state of the Hamiltonian (\ref{IHamXXZ}). In the 
framework of the algebraic Bethe Ansatz \cite{FadST79} it can be described
in terms of the generalised creation and annihilation operators which can be obtained as 
elements of the quantum monodromy matrix. This matrix is completely defined by the $R$-matrix
of the model which, for the XXZ chain, is the usual trigonometric solution of the Yang-Baxter equation,
acting in the tensor product of two auxiliary spaces $V_1
\otimes V_2$, $V_i=\mathbb{C}^2$:
%
%
%
%
%
\be{RMAtr}
R(\la)=\left(\ba{cccc}1&0&0&0\\
0&\frac{\sinh\la}{\sinh(\la-i\zeta)}&-\frac{i\sin\zeta}{\sinh(\la-i\zeta)}&0\\
0&-\frac{i\sin\zeta}{\sinh(\la-i\zeta)}&\frac{\sinh\la}{\sinh(\la-i\zeta)}&0\\
0&0&0&1\ea\right),\ee
the parameter $\zeta$ being related to the anisotropy parameter\footnote[2]{We give all the formulae 
in this paper for the critical regime ($-1<\Delta<1$) of the XXZ chain, but it will be clear
that similar 
computations can be done for the massive regime ($\Delta>1$) and for the XXX chain ($\Delta=1$).}
\[\Delta=\cos\zeta.\]
 The monodromy matrix can be constructed  as a
product of $R$-matrices and can be written as a $2\times 2$ matrix in the auxiliary space:
\be{Defmon}
T(\la)=R_{ 0 M}(\la-\xi_N+i\frac\zeta 2)\dots R_{ 0 1}(\la-\xi_1+i\frac\zeta 2)=
                                        \left(\ba{cc} A(\la)& B(\la)\\
                                                      C(\la)& D(\la)\ea\right)_{[0]}.
\ee
Here $\{\xi\}$ is a set of arbitrary inhomogeneity parameters and the $R$-matrices $R_{0 k}$ act
in the tensor product of the auxiliary space $V_0$ and the local quantum space $\mathcal{H}_k$.
The operator entries of the monodromy matrix $A$, $B$, $C$ and $D$ act in the same quantum 
space $\mathcal{H}$ as the Hamiltonian of the XXZ chain. The commutation relations of these operators can
be obtained from the Yang-Baxter equation:
\be{commutation}
R_{12}(\lambda- \mu)\ T_1(\lambda)\ T_2(\mu)
= T_2(\mu)\ T_1(\lambda)\ R_{12}(\lambda- \mu).
\ee
 From this relation one can easily see also that traces
of the monodromy matrix taken in the auxiliary space ({\it transfer matrices})
commute for any values of the spectral
parameters,
\be{transmat}
[A(\la)+D(\la),A(\mu)+D(\mu)]=0.
\ee 
The Hamiltonian of the XXZ model can be reconstructed from the transfer matrix 
in the homogeneous limit when all the inhomogeneity parameters are equal: $\xi_k=0$,
\begin{equation}
H=c\left.\pd{\la}\log(A(\la)+D(\la))\right|_{\la=0}+\mathrm{const}.
\end{equation}
It means in particular
 that the Hamiltonian commutes with the transfer matrix for any value of the spectral
parameter,
\[  [H,A(\la)+D(\la)]=0,\]
and the eigenstates of the transfer matrix (for arbitrary $\la$) are eigenstates of the 
Hamiltonian. 

To construct the eigenstates of the transfer matrix one can use the operators $B(\la)$
as creation operator (and operators $C(\la)$ as annihilation operators). This is possible if there
exists a reference state $\ket{0}$ which is an eigenstate of the operators $A(\la)$ and $D(\la)$ 
and is annihilated by the operators $C(\la)$ for any value of $\la$.
%
For the XXZ chain such a state exists and it is the ferromagnetic state with all the spins up.
Now other eigenstates can be constructed by the action of operators $B$ on this ferromagnetic
state. More precisely, using the Yang-Baxter algebra (\ref{commutation}), one can show that 
the eigenstates of the transfer matrix (and hence of the Hamiltonian in the homogeneous case) can be
constructed in the form:
\be{Bethestate}
\ket{\psi}=B(\la_1)\dots B(\la_N)\ket{0},\ee
where the spectral parameters $\{\la_j\}$ satisfy the {\it Bethe equations},
\be{ABABE}
\prod_{m=1}^M\frac{\sinh(\lambda_j-\xi_m+i\frac\zeta 2)}{\sinh(\lambda_j-\xi_m-i\frac\zeta 2)}
\cdot\prod_{k=1\atop{k\ne
j}}^{N}  \frac{\sinh(\lambda_j-\lambda_k-i\zeta)}
{\sinh(\lambda_j-\lambda_k+i\zeta)}=1,
\qquad j=1,\dots, N.
\ee
The Bethe equations are very difficult to solve for a finite chain.
 However it can be shown that the ground state 
of the XXZ chain is described by this procedure in the homogeneous case with $N=\frac M2$
 and can be specified by a special choice of 
integers in the logarithmic form of the Bethe equations:
\be{logBAE}
p_0(\la_j)-\frac 1M  
\sul_{k=1}^{ M/2}\phi(\la_j-\la_k)
=-\frac \pi 2-\frac{\pi}M +\frac{2\pi j}M,\qquad j=1,\dots, \frac M2,
\ee
where the  bare momentum $p_0(\la)$ and the ``scattering phase'' $\phi(\la)$ are defined as
\[p_0(\la)=i\log\frac{\sinh(\lambda+i\frac\zeta 2)}{\sinh(\lambda-i\frac\zeta 2)}\qquad
\phi(\la)=i\log\frac{\sinh(\lambda+i\zeta )}{\sinh(\lambda-i\zeta )}.\]
This state is the ground state of the XXZ chain in the homogeneous case, but even for the
inhomogeneous model one can define a Bethe state with this choice of integers in the right 
hand side of the logarithmic Bethe equations. 
In the calculation of the correlation functions 
it will be convenient to consider such a state first
and to take the homogeneous limit only in the final result.

Even for the ground state there is no way in a generic situation  to solve the Bethe equations
explicitly. However, in the thermodynamic limit $M\rightarrow\infty$, 
this state can be described in a very simple way in terms of the {\it density of rapidities}.
More precisely: in the thermodynamic limit, for any smooth bounded function $f(\la)$, any
sum over the Bethe roots corresponding to the ground state can be replaced by the following 
integral:
\be{sum-int}
\frac 1M\sul_{j=1}^{M/2} f(\la_j)=\int\limits_{-\infty}^{\infty}d\la \rho(\la)f(\la)+o(\frac 1M),
\ee
where the density function $\rho(\la)$ can be obtained from a simple 
integral equation which replaces
in the thermodynamic limit the Bethe equations,
\be{LiebE}
\rho(\la)+\int\limits_{-\infty}^{\infty}d\mu\, \rho(\mu)K(\la-\mu)=\frac 1{2\pi}p_0'(\la),
\ee
 where the kernel $K(\la)$ is a derivative of the ``scattering phase''
\be{kernK}
K(\la)=\frac{1}{2\pi}\phi'(\la)=\frac 1{2\pi}\frac{\sin(2\zeta)}
{\sinh(\la+i\zeta)\sinh(\la-i\zeta)}.
\ee 
This equation can be easily solved by Fourier transform,
\be{density}\rho(\la)=\frac{1}{2\zeta \cosh(\frac\pi\zeta\la)}.\ee
This information about the ground state is sufficient for the calculation of the
correlation functions in the thermodynamic limit.

The first problem which arises when one tries to calculate ground state mean values of 
products of local operators in the framework of the algebraic Bethe Ansatz
 is the fact that creation (annihilation) operators $B$ ($C$) are
non local and do not permit a simple expansion in terms of spin operators. Thus it is
 rather difficult to establish commutation relations between these two types of objects. 
In our paper \cite{KitMT99} we proposed a way to solve this problem by expressing the
local operators in terms of the monodromy matrix elements (hence solving the quantum inverse
problem). The expression for the local elementary $2\times 2$ 
matrices $E_m^{\epsilon'_m \epsilon_m}$
 with only one non-zero entry
\[E_{j k}^{\epsilon' \epsilon}=\delta_{j\epsilon'}\delta_{k \epsilon},\] 
can be written in a very simple form,
\be{InvPro}
E^{\epsilon'_m,\epsilon_m}_m=\prod_{k=1}^{m-1}
\Bigl(A+D\Bigr)(\xi_k-i\frac\zeta 2)
T_{\epsilon_m,\epsilon'_m}(\xi_m-i\frac\zeta 2)\prod_{k=1}^{m}
\Bigl(A+D\Bigr)^{-1}(\xi_k-i\frac\zeta 2).
\ee
It is easy to see that every local operator is expressed as a monodromy matrix element
dressed with transfer matrices. In particular the operator $\sigma^+_m$ is expressed as the 
dressed
 operator $C(\xi_m-i\frac\zeta 2)$, $\sigma^-_m$ as $B(\xi_m-i\frac\zeta 2)$ 
and $\sigma_m^z$ as $A(\xi_m-i\frac\zeta 2)-D(\xi_m-i\frac\zeta 2)$. 
It is important to mention that this solution is very convenient  to calculate the ground
state mean values as the ground state is an eigenstate of the transfer matrix,
\[(A(\xi_k-i\frac\zeta 2)+D(\xi_k-
i\frac\zeta 2))\ket{\psi_g}=\pl_{j=1}^{M/2}\frac{\sinh(\la_j-\xi_k-i\frac\zeta 2)}
{\sinh(\la_j-\xi_k+i\frac\zeta 2)}\ket{\psi_g}.\]
Now the correlation functions can be expressed only in terms of the monodromy matrix elements.
The first type of object we can consider using this approach are the ``elementary blocks'' i.e.
the ground state mean values of products of the elementary local matrices in $m$ consecutive sites,
\be{elblocks}
F_m(\{\epsilon_j,\epsilon'_j\})=\bra{\psi_g}\pl_{j=1}^m E^{\epsilon'_j,\epsilon_j}_j\ket{\psi_g}.
\ee 
Using the solution of the quantum inverse problem such quantities can be expressed only in 
terms of the monodromy matrix elements,
\be{elblocksT}
F_m(\{\epsilon_j,\epsilon'_j\})=\left(\pl_{k=1}^m\pl_{j=1}^{M/2}\frac{\sinh(\la_j-\xi_k+i\frac\zeta 2)}
{\sinh(\la_j-\xi_k-i\frac\zeta 2)}\right)
\frac{\bra{0}\pl_{j=1}^{M/2}C(\la_j)\,\pl_{k=1}^m T_{\epsilon_k,\epsilon'_k}(\xi_k-i\frac\zeta 2)\,
\pl_{j=1}^{M/2}B(\la_j)\ket{0}}{\bra{0}\pl_{j=1}^{M/2}C(\la_j)\,
\pl_{j=1}^{M/2}B(\la_j)\ket{0}},
\ee 
(it is important to mention that operators $B(\la)$ are not normalised and thus to obtain the mean value,
one should divide  the r.h.s. by the norms of the Bethe vector corresponding to the ground state).
Now we can use the Yang-Baxter algebra i.e. the commutation relations between the monodromy
matrix elements. In particular we can act with the monodromy matrix elements on the dual Bethe
state constructed by the actions of operators $C(\la)$. 

It is easy to see that after acting with all the operators in (\ref{elblocksT}) on the dual
Bethe state one obtains again a sum of states constructed by the action of the operators
$C(\la)$ on the dual ferromagnetic state (but the spectral parameters do not satisfy any
more the Bethe equations). Then, using 
the Gaudin-Korepin formula for the norm of the Bethe states \cite{GauMW81,Gau83L,Kor82},
%
%
 and the determinant representation for the scalar products of a Bethe state with an arbitrary 
state
\cite{Sla89,KitMT99}, 
%
one can express the  correlation functions (\ref{elblocks}) as sums of determinants.
Now we have an explicit formula for the elementary blocks in term of the ground state solution of the
Bethe equations. In the thermodynamic limit this sum can be simplified as all the sums over the
Bethe roots can be replaced by integrals with density. The final result for any elementary 
block can be written as multiple integrals
\be{resultEB}
F_m(\{\epsilon_j,\epsilon'_j\})=\frac 1{\pl_{j>k}\sinh(\xi_j-\xi_k)}
\int\limits_{-\infty}^{\infty}d\la_1\dots
\int\limits_{-\infty}^{\infty}d\la_m\, 
\mathcal{F}(\{\la_k\},\{\xi_j,\epsilon_j,\epsilon'_j\})\det_m 
S(\{\la_j\},\{\xi_k\}).
\ee
Here the $m\times m$ matrix  $S$ does not depend on the choice of local operator and is defined uniquely
by the ground state. Its elements can be written in terms of the density function (\ref{density})
\be{densityM}
S_{j k}=\rho(\la_j-\xi_k).
\ee
 The algebraic part $
\mathcal{F}(\{\la_k\},\{\xi_j,\epsilon_j,\epsilon'_j\})$ arises from the commutation 
relation of the monodromy matrix elements and does not depend on the ground state.
The expression for this function for the most general case can be found in the paper \cite{KitMT00}.
Here for the calculation of the two-point functions we need only some particular blocks
and we will give an explicit expression for the corresponding algebraic parts 
in the next sections.

Once the elementary blocks are calculated, any correlation function can be written
as a sum of such multiple integrals. However the sums which  appear for the two-point
functions contain  a number of terms  that grows exponentially with the  distance 
and in such a form cannot 
be used for the asymptotic analysis.
For  example, using the solution of the quantum inverse problem for the correlation functions of the third components of spin $g_{z z}(m)$, 
we obtain the following expression:
\begin{align}
g_{z z}(m)=&\left(\pl_{k=1}^{m+1}\pl_{j=1}^{M/2}\frac{\sinh(\la_j-\xi_k+i\frac\zeta 2)}
{\sinh(\la_j-\xi_k-i\frac\zeta 2)}\right)\bra{\psi_g}(A(\xi_1-i\frac\zeta 2)-D(\xi_1-i\frac\zeta 2))
\nonumber\\
&\times
\left(\pl_{j=2}^m(A(\xi_j-i\frac\zeta 2)+D(\xi_j-i\frac\zeta 2))\right)(
A(\xi_{m+1}-i\frac\zeta 2)-D(\xi_{m+1}-i\frac\zeta 2))\ket{\psi_g}.
\label{zzInvprob}
\end{align}
It is easy to see that this function can be rewritten as a sum of $2^{m+1}$ elementary
blocks. Similarly, for the function $g_{+-}(m)$, one has
\begin{align}
g_{+ -}(m)=&\left(\pl_{k=1}^{m+1}\pl_{j=1}^{M/2}\frac{\sinh(\la_j-\xi_k+i\frac\zeta 2)}
{\sinh(\la_j-\xi_k-i\frac\zeta 2)}\right)\nonumber\\
&\times\bra{\psi_g}C(\xi_1-i\frac\zeta 2)
\left(\pl_{j=2}^m(A(\xi_j-i\frac\zeta 2)+D(\xi_j-i\frac\zeta 2))\right)
B(\xi_{m+1}-i\frac\zeta 2)\ket{\psi_g}.
\label{+-Invprob}
\end{align}
and it can be written as a sum of $2^{m-1}$ elementary blocks. The main goal of this paper 
is to obtain a manageable expression for these correlation functions from the multiple
integral representation for the elementary blocks.

\section{Generating function}

In this section we consider the correlation function of the  third components of spin
$g_{z z}(m)$. To calculate this function it is convenient to introduce a new object:
 the {\it generating function}
\be{GFdef}
\mathcal{Q}_m(\b)\equiv \bra{\psi_g}\exp\{\b Q_{1,m}\}\ket{\psi_g}, 
\qquad Q_{1,m}=\sul_{j=1}^m \frac 12(1-\s_j^z).
\ee 
To obtain the two-point function  from the generating function one 
should take its second derivative on $\b$ and second lattice derivative on $m$ 
 \be{GF->zz}
g_{z z}(m)= \left.\left( 2\mathcal{D}^2_m\frac{\partial^2}{\partial\b^2}-
4\mathcal{D}_m\frac{\partial}{\partial\b}+1\right)\mathcal{Q}_m(\b)\right|_{\b=0},
\ee
where we used the standard definition of the first and second lattice derivatives
\[\mathcal{D}_m f(m)=f(m+1)-f(m),\qquad \mathcal{D}^2_m f(m)=f(m+1)-2f(m)+f(m-1).\]
Using the solution of the quantum inverse problem we can rewrite this 
quantity in  terms of the monodromy matrix elements: 
\be{GFAD}
\mathcal{Q}_m(\b)=\left(\pl_{k=1}^m\pl_{j=1}^{M/2}\frac{\sinh(\la_j-\xi_k+i\frac\zeta 2)}
{\sinh(\la_j-\xi_k-i\frac\zeta 2)}\right)
\bra{\psi_g}
\left(\pl_{j=1}^m(A(\xi_j-i\frac\zeta 2)+e^\b D(\xi_j-i\frac\zeta 2))\right)\ket{\psi_g}.
\ee
 The generating function can be expressed as a sum 
of $2^m$ elementary blocks containing only diagonal elements of the monodromy matrix 
(operators $A(\xi-i\frac\zeta 2)$ and $D(\xi-i\frac\zeta 2)$). 
For such elementary 
blocks the algebraic part can be written in a factorised form, as a product of 
``two-particle contributions'', 
\begin{align}
&\left(\pl_{k=1}^m\pl_{j=1}^{M/2}\frac{\sinh(\la_j-\xi_k+i\frac\zeta 2)}
{\sinh(\la_j-\xi_k-i\frac\zeta2)}\right)\bra{\psi_g}T_{a_1 a_1}(\xi_1-i\frac\zeta 2)\dots
T_{a_m a_m}(\xi_m-i\frac\zeta 2)\ket{\psi_g}\nonumber\\
=&\frac 1{\pl_{j<k}\sinh(\xi_j-\xi_k)}\int\limits_{-\infty}^{\infty}d\la_1\dots
\int\limits_{-\infty}^{\infty}d\la_m\det_m S(\{\la_k\},\{\xi_j\})\nonumber\\
\times&
\pl_{j>k}\frac{\sinh(\la_j-\xi_k+i\epsilon_j\zeta)
\sinh(\la_k-\xi_j-i\epsilon_k\zeta)}{\sinh(\la_j-\la_k+i(\epsilon_j+\epsilon_k)\zeta)}
,
\end{align}
where numbers $\epsilon_j=\pm\frac 12$ depend on the choice of the corresponding monodromy 
matrix elements $\epsilon_j=\frac 32-a_j$.

 It is clear that the elementary blocks with fixed number of operators of each
type have  a rather similar structure. Thus it is quite natural to put together the terms with the
same number of operators $D$ and to represent the generating function as power series on $e^\b$,
\begin{equation}
\mathcal{Q}_m(\b)= \sul_{s=0}^m e^{s\b } F_s(m).
\label{sum_gf1a}
\end{equation}
The main result of this paper is a simple and compact formula for the terms $F_s(m)$. 
This function can be expressed in terms of elementary blocks,
\begin{equation}
F_s(m)= \left(\pl_{k=1}^m\pl_{j=1}^{M/2}\frac{\sinh(\la_j-\xi_k+i\frac\zeta 2)}
{\sinh(\la_j-\xi_k-i\frac\zeta 2)}\right)\sul_{a_1+\dots +a_m-m=s} \bra{\psi_g}
T_{a_1 a_1}(\xi_1-i\frac\zeta 2)\dots
T_{a_m a_m}(\xi_m-i\frac\zeta 2)\ket{\psi_g},
\end{equation}
and hence the  sums over all possible positions of operators $D$ and $A$ also can be written as 
multiple integrals
\begin{equation}
F_s(m)= \pl_{j<k}\frac 1{\sinh(\xi_j-\xi_k)}\int\limits_{-\infty}^{\infty}d\la_1\dots
\int\limits_{-\infty}^{\infty}d\la_m \,\,G_s(m,\{\la_j\}|\{\xi_k\})
\det_m S(\{\la_j\},\{\xi_k\}),
\end{equation}
 where  the function $G_s(m,\{\la_j\}|\{\xi_k\})$ under the integrals is a sum 
over the permutations $\sigma$ of the set $1,\dots,m$
\begin{equation}
G_s(m,\{\la_j\}|\{\xi_k\})=\!\frac 1{s!(m-s)!}\!
\sul_{\sigma}(-1)^{[\sigma]}\pl_{j>k}\frac{\sinh(\la_{\s(j)}-\xi_k+
i\epsilon_{\s(j)}\zeta)
\sinh(\la_{\s(k)}-\xi_j-i\epsilon_{\s(k)}\zeta)}{\sinh(\la_{\s(j)}-\la_{\s(k)}+i(\epsilon_{\s(j)}+
\epsilon_{\s(k)})\zeta)},
\label{permsum}
\end{equation}
where $(-1)^{[\sigma]}$ is the sign of the permutation $\sigma$, 
the  factorials in the denominator arise from the additional summations over
permutations of the variables of the same type  ( with $\epsilon_j=\frac 12$ or $\epsilon_j=-\frac 12$).
We set for $j\le s$, $\epsilon_j=-\frac 12$ and  for $j> s$, $\epsilon_j=\frac 12$. 
We use here a very
important property of the multiple integral representation, namely, the fact that
the determinant of densities does not depend on the choice of local operators.

The function $G_s(m,\{\la_j\}|\{\xi_k\})$ is a rational function of $e^{\la_j}$ and $e^{\xi_k}$. 
It is by definition skew-symmetric under the permutations of $\la_1\dots,\la_m$.
The first property of this function which we need for our computation is its symmetry on the variables
$\xi_j$.
\begin{lemma}
\label{symmetry}
The function $G_s(m,\{\la_j\}|\{\xi_k\})$ defined by (\ref{permsum}) is symmetric under the
permutations  of the variables $\xi_1,\xi_2,\dots ,\xi_m$.  
\end{lemma}
{\bf Proof:\,\,} To prove this lemma it is sufficient to show for any $k$ that
\[G_s(m,\{\la_j\}|\xi_1,\dots,\xi_k,\xi_{k+1},\dots,\xi_m)-
G_s(m,\{\la_j\}|\xi_1,\dots,\xi_{k+1},\xi_{k},\dots,\xi_m)=0\]
Consider first this difference for corresponding monomials in (\ref{permsum}) and note 
that  two terms differ only in one ``two-particle contribution'':
\begin{align*}
&\pl_{j>l\atop{l\neq k}}\frac{\sinh(\la_j-\xi_l+i\epsilon_j\zeta)
\sinh(\la_l-\xi_j-i\epsilon_l\zeta)}{\sinh(\la_j-\la_l+i(\epsilon_j+\epsilon_l)\zeta)}\pl_{j=k+2}^m
\frac{\sinh(\la_j-\xi_k+i\epsilon_j\zeta)
\sinh(\la_k-\xi_j-i\epsilon_k\zeta)}{\sinh(\la_j-\la_k+i(\epsilon_j+\epsilon_k)\zeta)}\\
&\times\left(
\frac{\sinh(\la_{k+1}-\xi_k+i\epsilon_{k+1}\zeta)
\sinh(\la_k-\xi_{k+1}-i\epsilon_k\zeta)}
{\sinh(\la_{k+1}-\la_k+i(\epsilon_{k+1}+\epsilon_k)\zeta)}\right.\\
&\hphantom{\frac{\sinh(\la_j-\xi_l+i\epsilon_j\zeta)
\sinh(\la_l-\xi_j-i\epsilon_l\zeta)}{\sinh(\la_j-\la_l+i(\epsilon_j+\epsilon_l)\zeta)}}
\left.-\frac{\sinh(\la_{k+1}-\xi_{k+1}+i\epsilon_{k+1}\zeta)
\sinh(\la_k-\xi_{k}-i\epsilon_k\zeta)}{\sinh(\la_{k+1}-\la_k+i(\epsilon_{k+1}+\epsilon_k)\zeta)}\right)\\
=&
\sinh(\xi_{k}-\xi_{k+1})\pl_{j>l\atop{l\neq k}}\frac{\sinh(\la_j-\xi_l+i\epsilon_j\zeta)
\sinh(\la_l-\xi_j-i\epsilon_l\zeta)}{\sinh(\la_j-\la_l+i(\epsilon_j+\epsilon_l)\zeta)}\\
&\hphantom{\frac{\sinh(\la_j-\xi_l+i\epsilon_j\zeta)
\sinh(\la_l-\xi_j-i\epsilon_l\zeta)}{\sinh(\la_j-\la_l+i(\epsilon_j+\epsilon_l)\zeta)}}
\times\pl_{j=k+2}^m
\frac{\sinh(\la_j-\xi_k+i\epsilon_j\zeta)
\sinh(\la_k-\xi_j-i\epsilon_k\zeta)}{\sinh(\la_j-\la_k+i(\epsilon_j+\epsilon_k)\zeta)}.
\end{align*}
Consider now the same term for the following  permutation  of the set $\la_1,\dots\la_m$:
\[\{\la_1,\dots,\la_{k-1},\la_{k+1},\la_k,\la_{k+2},\dots,\la_m\}.\] 
It is easy to see 
that for this permutation we obtain exactly the same contribution with an opposite
sign (as the sign of this permutation is $-1$) and the sum of these two contributions 
is zero.
Now we should take a sum of such monomials
 over all possible permutations of the set $\la_1,\dots,\la_m$, but as this sum can
be split into such pairs with permuted $\la_{\sigma(k)}$ and 
$\la_{\sigma(k+1)}$ we immediately obtain that this sum is  zero.
\qed
Using this symmetry we can obtain  recursion relations for the function  
$G_s(m,\{\la_j\}|\{\xi_k\})$ in the points $\la_j=\xi_k-i\frac\zeta 2$. 
It is more convenient now to extract the
common denominator and to  consider the  function $\tilde{G}_s(m,\{\la_j\}|\{\xi_k\})$:
\begin{align*}
G_s(m,\{\la_j\}|\{\xi_k\})&=\frac 1{s!(m-s)!}
\pl_{j>k}\frac{\sinh(\la_j-\la_k)}{\sinh(\la_j-\la_k+
i(\epsilon_j+\epsilon_k)\zeta)\sinh(\la_j-\la_k-
i(\epsilon_j+\epsilon_k)\zeta)}\\&\times \tilde{G}_s(m,\{\la_j\}|\{\xi_k\})
\end{align*}
Directly from this definition we can establish two evident lemmas:
\begin{lemma}
The function $e^{(m-1)\la_j}\tilde{G}_s(m,\{\la_j\}|\{\xi_k\})$ is a polynomial function 
of $\e^{2\la_j}$
of degree $m-1$. 
\end{lemma}
\begin{lemma}
\[\tilde{G}_0(1,\la_1|\xi_1)=\tilde{G}_1(1,\la_1|\xi_1)=1.\]
\end{lemma}
These lemmas mean that if we obtain recurrence relations  for this function
in the points   $\la_j=\xi_k-i\frac\zeta 2$ then they will define it completely
(as it is a polynomial of degree $m-1$
defined in $m$ points). The recursion relation for this function can be written in 
the following form
\begin{lemma}
$\tilde{G}_s(m,\{\la_j\}|\{\xi_k\})$ satisfies the following recursion relations,
\begin{align}
 \tilde{G}_s(m,\{\la_l\}&|\{\xi_k\})\left.\vphantom{\pl_{j=1}}
\right|_{\la_j=\xi_k-i\frac \zeta 2}=\pl_{a=1\atop{a\neq k}}^m
\sinh(\la_j-\xi_a-i\frac \zeta2)\pl_{a\neq j}
\sinh(\la_a-\xi_k-i\frac\zeta 2)\nonumber\\
&\times \tilde{G}_{s}(m-1,\la_1,\dots,\la_{j-1},\la_{j+1},\dots,\la_m|
\xi_1,\dots,\xi_{k-1},\xi_{k+1},\dots\xi_m), 
\quad 
 \epsilon_j=\frac 12 \label{recurs1}\\
\tilde{G}_s(m,\{\la_l\}&|\{\xi_k\})\left.\vphantom{\pl_{j=1}}
\right|_{\la_j=\xi_k-i\frac \zeta 2}=\pl_{a=1\atop{a\neq k}}^m
\sinh(\la_j-\xi_a-i\frac \zeta2)\pl_{a\neq j}
\sinh(\la_a-\xi_k-i\frac\zeta 2)\nonumber\\
&\times \tilde{G}_{s-1}(m-1,\la_1,\dots,\la_{j-1},\la_{j+1},\dots,\la_m|
\xi_1,\dots,\xi_{k-1},\xi_{k+1},\dots\xi_m), 
\quad 
 \epsilon_j=-\frac 12\label{recurs2}
\end{align}
\end{lemma}
{\bf Proof:}\,\, From the Lemma \ref{symmetry} it is clear that it is sufficient to prove 
these recursion relations only for $\la_j=\xi_1-i\frac\zeta 2$ for $\epsilon_j=\frac 12 $
and for $\la_j=\xi_m-i\frac\zeta 2$ for $\epsilon_j=-\frac 12$. 
It is easy to see that only terms with $\sigma(j)=1$ ($\sigma(j)=m$ ) will survive
in the sum over permutations (\ref{permsum}). We obtain a sum over permutations of the set
$\{\la_1,\dots,\la_{j-1},\la_{j+1},\dots,\la_m\}$ and using the definition of the
function $\tilde{G}_s(m,\{\la_l\}|\{\xi_k\})$ we obtain immediately the  recursion
relations (\ref{recurs1}), (\ref{recurs2}).\qed
These recursion relations define completely the function $\tilde{G}_s(m,\{\la_l\}|\{\xi_k\})$. It means that 
if we can find a function satisfying  Lemmas 3.1-3.4 
it is the function $\tilde{G}_s(m,\{\la_l\}|\{\xi_k\})$
 we need. Such properties  (without dependence on $s$)  were 
established for the first time by Korepin in \cite{Kor82} for the partition function of the six vertex model with domain 
wall boundary conditions. In fact the recursion relations obtained here are exactly the 
same as the Korepin's ones. The (unique) solution for these relations, found by Izergin in \cite{Ize87} 
(which does not depend on $s$), 
satisfies the Lemmas 3.1-3.4 
for any value of $s$ and thus gives the function $\tilde{G}_s(m,\{\la_l\}|\{\xi_k\})$:
\begin{theorem}[Izergin 87]
The only function satisfying lemmas 3.1-3.4 is  
\[\tilde{G}_s(m,\{\la_l\}|\{\xi_k\})=\frac 1{\sin^m\zeta}Z_m(\{\la_l\}|\{\xi_k\}),\]
 where  $Z_m(\{\la\}|\{\xi\})$ is the partition function of the inhomogeneous 
six vertex model with  domain wall boundary conditions, which
can be written in the following form,
\begin{align}
Z_m(\{\la_l\}|\{\xi_k\})=&\frac {\pl_{j=1}^m\pl_{k=1}^m\sinh(\la_j-\xi_k+i\frac \zeta2)
\sinh(\la_j-\xi_k-i\frac \zeta2)}{\pl_{j>k}\sinh(\la_j-\la_k)\sinh(\xi_k-\xi_j)}\det_m 
\mathcal{M}(\{\la_l\}|\{\xi_k\}),\\
\mathcal{M}_{j k}=&\frac {\sin\zeta}{\sinh(\la_j-\xi_k+i\frac \zeta2)
\sinh(\la_j-\xi_k-i\frac \zeta2)}.
\end{align}
\end{theorem}  
{\bf Proof:} \,\, To prove this theorem one needs just to verify that the function
$Z_m(\{\la\}|\{\xi\})$ satisfies all 
the Lemmas 3.1-3.4 
for any value of $s$. Then, as it is a polynomial of degree $m-1$ coinciding with another 
polynomial of the same degree in $m$ points, one can conclude that these two functions
are equal.\qed

The fact that the function $\tilde{G}_s(m,\{\la_l\}|\{\xi_k\})$ does not depend on $s$ is a very important
peculiarity of this re-summation technique.
Hence, the generating function can be written as a sum of $m+1$ multiple integrals of the 
following form,
\begin{align}
F_s(m)=&\frac 1{s!(m-s)!\sin^m\zeta}\pl_{j<k}
\frac 1{\sinh(\xi_j-\xi_k)}\int\limits_{-\infty}^{\infty}\! d\la_1\dots\!\!
\int\limits_{-\infty}^{\infty}\! d\la_m\, \det_m S(\{\la\},\{\xi\}) 
\nonumber\\ 
&\times
\Theta_m^s(\la_1,\dots,\la_m)\,\,\, Z_m(\{\la\}|\{\xi\})\, ,\label{resum_gff}
\end{align}
where the only factor depending on $s$ which we denote $\Theta_m^s(\{\la\})$
\begin{align}
\Theta_m^s(\la_1,\dots,\la_m)=&\pl_{k=1}^s\pl_{j=s+1}^m
\frac 1{\sinh(\la_j-\la_k)}\pl_{m\ge j>k> s}
\frac{\sinh(\la_j-\la_k)}{\sinh(\la_j-\la_k+i\zeta)\sinh(\la_j-\la_k-i\zeta)}\nonumber\\ 
&\times
 \pl_{s\ge j>k\ge 1}
\frac{\sinh(\la_j-\la_k)}{\sinh(\la_j-\la_k+i\zeta)\sinh(\la_j-\la_k-i\zeta)
}
\end{align}
will appear in all our results.

Thus we reduced the number of terms from  exponential to polynomial order. This representation
is also interesting because of its unexpected relation with 
 the partition function of the corresponding inhomogeneous six-vertex model
with domain wall boundary conditions.
We also suppose that this representation can be convenient for the asymptotic analysis of the 
two-point function $g_{z z}(m)$.

It is also very important to mention that for this re-summation  we manipulated only the algebraic
part of the expression for the elementary blocks and hence it can be done in a similar way for 
the XXZ spin chain in a magnetic field. The result in this case has a slightly more complicated 
deteminant of densities  and different integration contours but the algebraic part is the same and
contains the partition function $ Z_m(\{\la\}|\{\xi\})$. 

A rough asymptotic analysis of the generating function can be performed using a 
modification of the saddle point 
technique introduced in \cite{KitMST02a} for the  emptiness formation probability.
It shows that there is no gaussian contribution to this correlation function and that the main order 
can be written as $C\exp(\frac {m\b} 2)$, as it should be, 
but it is not sufficient to describe the power-like  behaviour 
of the two-point function.  The interesting pecularity of this saddle point analysis is the fact that the
``saddle point density'' here coincides with the ground state density $\rho(\la)$.    

\section{Two-point functions}

A similar re-summation can be done for the two-point functions $g_{+-}(m)$ (or for $g_{xx}(m)$ and
$g_{yy}(m)$) and $g_{zz}(m)$. The function $g_{+-}(m)$ can be written
in the following form in terms of the monodromy matrix elements:
\begin{align}
g_{+ -}(m)=&\left(\pl_{k=1}^{m+1}\pl_{j=1}^{M/2}\frac{\sinh(\la_j-\xi_k+i\frac\zeta 2)}
{\sinh(\la_j-\xi_k-i\frac\zeta 2)}\right)\nonumber\\
&\times\bra{\psi_g}C(\xi_1-i\frac\zeta 2)
\left(\pl_{j=2}^m(A(\xi_j-i\frac\zeta 2)+D(\xi_j-i\frac\zeta 2))\right)B(\xi_{m+1}-i\frac\zeta 2)
\ket{\psi_g}.
\label{+-Invprob2}
\end{align}
To calculate this function one should sum up $2^{m-1}$ elementary blocks of the following form
\begin{align*}
g_{+ -}(m)=&\left(\pl_{k=1}^{m+1}\pl_{j=1}^{M/2}\frac{\sinh(\la_j-\xi_k+i\frac\zeta 2)}
{\sinh(\la_j-\xi_k-i\frac\zeta 2)}\right)\nonumber\\
&\times\sul_{a_j=1,2}\bra{\psi_g}C(\xi_1-i\frac\zeta 2)
\left(\pl_{j=2}^mT_{a_j a_j}(\xi_j-i\frac\zeta 2)\right)B(\xi_{m+1}-i\frac\zeta 2)
\ket{\psi_g},
\end{align*}
which can be written as multiple integrals in a factorised form
\begin{align}
&\left(\pl_{k=1}^{m+1}\pl_{j=1}^{M/2}\frac{\sinh(\la_j-\xi_k+i\frac\zeta 2)}
{\sinh(\la_j-\xi_k-i\frac\zeta2)}\right)\nonumber\\
\times&\bra{\psi_g}C(\xi_1-i\frac\zeta 2)T_{a_2 a_2}(\xi_2-i\frac\zeta 2)\dots
T_{a_m a_m}(\xi_m-i\frac\zeta 2)B(\xi_{m+1}-i\frac\zeta 2)\ket{\psi_g}\nonumber\\
=&\frac 1{\pl_{k>j\ge 1}^{m+1}\sinh(\xi_j-\xi_k)}\int\limits_{-\infty}^{\infty}d\la_2\dots
\int\limits_{-\infty}^{\infty}d\la_m\int\limits_{-\infty}^{\infty}d\la_+
\int\limits_{-\infty}^{\infty}d\la_-
\det_{m+1} S(\{\la_2,\dots,\la_m,\la_+,\la_-\},\{\xi_j\})\nonumber\\\times&
\left(\pl_{j>k\ge 2}^m \frac{\sinh(\la_j-\xi_k+i\epsilon_j\zeta)
\sinh(\la_k-\xi_j-i\epsilon_k\zeta)}{\sinh(\la_j-\la_k+i(\epsilon_j+\epsilon_k)\zeta)}\right)
\frac{\sinh(\la_+ -\xi_1+i\frac\zeta 2)
\sinh(\la_- -\xi_1-i\frac\zeta 2)}{\sinh(\la_+ -\la_-)}
 \nonumber\\
\times&\pl_{k=2}^m \frac{\sinh(\la_- -\xi_k-i\frac\zeta 2)
\sinh(\la_k-\xi_1+i\epsilon_k\zeta)}{\sinh(\la_- -\la_k+i(\epsilon_k-\frac 12)\zeta)}
 \nonumber\\
\times&\pl_{k=2}^m \frac{\sinh(\la_+ -\xi_k+i\frac\zeta 2)
\sinh(\la_k-\xi_{m+1}-i\epsilon_k\zeta)}{\sinh(\la_+ -\la_k+i(\epsilon_k+\frac 12)\zeta)}
.
\end{align}
The blocks with the same number of operators $D$ ($\epsilon_j=-\frac 12)$  can be  put together 
\begin{equation}
g_{+ -}(m)= \sul_{s=0}^{m-1}  \tilde{g}_{+-}(m,s) ,
\end{equation}
where
\begin{align}
\tilde{g}_{+-}&(m,s)= \left(\pl_{k=1}^{m+1}\pl_{j=1}^{M/2}\frac{\sinh(\la_j-\xi_k+i\frac\zeta 2)}
{\sinh(\la_j-\xi_k-i\frac\zeta 2)}\right)\nonumber\\
\times&\sul_{a_2+\dots +a_m-m+1=s} \bra{\psi_g}C(\xi_1-i\frac\zeta 2)
T_{a_2 a_2}(\xi_1-i\frac\zeta 2)\dots
T_{a_m a_m}(\xi_m-i\frac\zeta 2)B(\xi_{m+1}-i\frac\zeta 2)\ket{\psi_g}.
\end{align}
It is easy to see that in all the terms of this sum written as multiple integrals the
determinant of densities and {\it all the factors} containing $\la_+$ and $\la_-$ are the same.
Thus to obtain the corresponding algebraic part it is sufficient to take the sum  over all possible
permutations of $\la_2,\dots,\la_m$ of the product 
\[\pl_{j>k\ge 2}^m \frac{\sinh(\la_j-\xi_k+i\epsilon_j\zeta)
\sinh(\la_k-\xi_j-i\epsilon_k\zeta)}{\sinh(\la_j-\la_k+i(\epsilon_j+\epsilon_k)\zeta)}.\]
This sum has exactly the same form as the corresponding term for the generating function.
Using exactly the same arguments as in the previous section we obtain for the 
contributions to the two-point function:
\begin{align}
\tilde{g}_{+-}(m,s)=& \frac 1{s!(m-1-s)!\sin^{m-1}\zeta}\pl_{m+1\ge k>j\ge 1}
\frac 1{\sinh(\xi_j-\xi_k)}\int\limits_{-\infty}^{\infty}\! d\la_2\dots\!\!
\int\limits_{-\infty}^{\infty}\! d\la_m \int\limits_{-\infty}^{\infty}\! d\la_+
\int\limits_{-\infty}^{\infty}\! d\la_-
\nonumber\\
\times&\left(\pl_{k=2}^{s+1} \frac{\sinh(\la_- -\xi_k-i\frac\zeta 2)
\sinh(\la_k-\xi_1-i\frac\zeta 2)}{\sinh(\la_- -\la_k-i\zeta)}\right) \nonumber\\
\times&
\left(\pl_{k=s+2}^{m} \frac{\sinh(\la_- -\xi_k-i\frac\zeta 2)
\sinh(\la_k-\xi_1+i\frac\zeta 2)}{\sinh(\la_- -\la_k)}
\right) \nonumber\\
\times&\left(\pl_{k=2}^{s+1} \frac{\sinh(\la_+ -\xi_k+i\frac\zeta 2)
\sinh(\la_k-\xi_{m+1}+i\frac\zeta 2)}{\sinh(\la_+ -\la_k)}\right) \nonumber\\
\times&
\left(\pl_{k=s+2}^{m} \frac{\sinh(\la_+ -\xi_k+i\frac\zeta 2)
\sinh(\la_k-\xi_{m+1}-i\frac\zeta 2)}{\sinh(\la_+ -\la_k+i\zeta)}\right)
\nonumber\\ \times&
\,\,\frac{\sinh(\la_+ -\xi_1+i\frac\zeta 2)
\sinh(\la_- -\xi_1-i\frac\zeta 2)}{\sinh(\la_+ -\la_-)}\,\,\cdot \Theta_{m-1}^s(\la_2,\dots,\la_m)
 \nonumber\\ \times& Z_{m-1}(\{\la_2,\dots,\la_m\}|\{\xi_2,\dots,\xi_m\})
\det_{m+1} S(\{\la_2,\dots,\la_m,\la_+,\la_-\},\{\xi_1,
\dots,\xi_{m+1}\}).\label{resum+-f}
\end{align}

This result is slightly more complicated but very similar to the result for the generating function.

One should note that a very similar formula can be written for the correlation function $g_{z z}(m)$
directly (without any use of the generating function).
 This function can be written
in the following form in terms of the monodromy matrix elements:
\begin{align}
g_{z z}(m)=& -1+4\left(\pl_{k=1}^{m+1}\pl_{j=1}^{M/2}\frac{\sinh(\la_j-\xi_k+i\frac\zeta 2)}
{\sinh(\la_j-\xi_k-i\frac\zeta 2)}\right)\nonumber\\
&\times\bra{\psi_g}D(\xi_1-i\frac\zeta 2)
\left(\pl_{j=2}^m(A(\xi_j-i\frac\zeta 2)+D(\xi_j-i\frac\zeta 2))\right)D(\xi_{m+1}-i\frac\zeta 2)
\ket{\psi_g}.
\label{+-Invprob2a}
\end{align}
To calculate this function one should sum up $2^{m-1}$ elementary blocks of the following form,
\begin{align*}
g_{z z}(m)=&-1+4\left(\pl_{k=1}^{m+1}\pl_{j=1}^{M/2}\frac{\sinh(\la_j-\xi_k+i\frac\zeta 2)}
{\sinh(\la_j-\xi_k-i\frac\zeta 2)}\right)\nonumber\\
&\times\sul_{a_j=1,2}\bra{\psi_g}D(\xi_1-i\frac\zeta 2)
\left(\pl_{j=2}^mT_{a_j a_j}(\xi_j-i\frac\zeta 2)\right)D(\xi_{m+1}-i\frac\zeta 2)
\ket{\psi_g}.
\end{align*}

As usual we put together the blocks with the same number of operators $D$,
\begin{equation}
g_{z z}(m)= -1+4\sul_{s=0}^{m-1}  \tilde{g}_{DD}(m,s) ,
\end{equation}
where
\begin{align}
\tilde{g}_{DD}&(m,s)= \left(\pl_{k=1}^{m+1}\pl_{j=1}^{M/2}\frac{\sinh(\la_j-\xi_k+i\frac\zeta 2)}
{\sinh(\la_j-\xi_k-i\frac\zeta 2)}\right)\nonumber\\
\times&\sul_{a_2+\dots +a_m-m+1=s} \bra{\psi_g}D(\xi_1-i\frac\zeta 2)
T_{a_2 a_2}(\xi_1-i\frac\zeta 2)\dots
T_{a_m a_m}(\xi_m-i\frac\zeta 2)D(\xi_{m+1}-i\frac\zeta 2)\ket{\psi_g}.
\end{align}

Using exactly the same arguments as for the  function $g_{+-}(m)$ we obtain for the 
contributions to the two-point function:
\begin{align}
\tilde{g}_{DD}(m,s)=& \frac 1{s!(m-1-s)!\sin^{m-1}\zeta}\pl_{m+1\ge j>k\ge 1}
\frac 1{\sinh(\xi_j-\xi_k)}\int\limits_{-\infty}^{\infty}\! d\la_1
\int\limits_{-\infty}^{\infty}\! d\la_2\dots\!\!
\int\limits_{-\infty}^{\infty}\! d\la_m 
\int\limits_{-\infty}^{\infty}\! d\la_{m+1}
\nonumber\\ 
\times&\left(\pl_{k=2}^{s+1} \frac{\sinh(\la_1 -\xi_k+i\frac\zeta 2)
\sinh(\la_k-\xi_1-i\frac\zeta 2)}{\sinh(\la_k -\la_1-i\zeta)}\right) \nonumber\\
\times&
\left(\pl_{k=s+2}^{m} \frac{\sinh(\la_1 -\xi_k+i\frac\zeta 2)
\sinh(\la_k-\xi_1+i\frac\zeta 2)}{\sinh(\la_k -\la_1)}
\right) \nonumber\\
\times&\left(\pl_{k=2}^{s+1} \frac{\sinh(\la_{m+1} -\xi_k-i\frac\zeta 2)
\sinh(\la_k-\xi_{m+1}+i\frac\zeta 2)}{\sinh(\la_{m+1}-\la_{k}-i\zeta)}\right) \nonumber\\
\times&
\left(\pl_{k=s+2}^{m} \frac{\sinh(\la_{m+1} -\xi_k-i\frac\zeta 2)
\sinh(\la_k-\xi_{m+1}-i\frac\zeta 2)}{\sinh(\la_{m+1}-\la_{k})}\right)
\nonumber\\ \times&
\,\,\frac{\sinh(\la_{m+1} -\xi_1-i\frac\zeta 2)
\sinh(\la_1 -\xi_{m+1}+i\frac\zeta 2)}{\sinh(\la_{m+1} -\la_{1}-i\zeta)}
\,\,\cdot \Theta_{m-1}^s(\la_2,\dots,\la_m)
 \nonumber\\ \times& Z_{m-1}(\{\la_2,\dots,\la_m\}|\{\xi_2,\dots,\xi_m\})
\det_{m+1} S(\{\la_1,\la_2,\dots,\la_m,\la_{m+1}\},\{\xi_1,
\dots,\xi_{m+1}\}).
\end{align}

\section{Further re-summations}

 In this section we show how one can proceed to a complete
 re-summation of terms and reduce the result for the two-point correlation functions
 to only one term (written again as a multiple integral). 
To do it one should note that all the terms in the sum
(\ref{sum_gf1a}) have a very similar structure. Extracting the common denominator one can obtain the following sum for the generating function:
\begin{align}
\mathcal{Q}_m(\b)= \sul_{s=0}^m&\frac {e^{s\b}}{s!(m-s)!\sin^m\zeta}\pl_{j<k}
\frac 1{\sinh(\xi_j-\xi_k)}\int\limits_{-\infty}^{\infty}\! d\la_1\dots\!\!
\int\limits_{-\infty}^{\infty}\! d\la_m \nonumber\\
&\times
\frac{\pl_{k=1}^m\pl_{j=1}^m\sinh(\la_j-\xi_k+i\frac \zeta 2)\sinh(\la_j-\xi_k-i\frac \zeta 2)}
{\pl_{ j> k}\sinh(\la_j-\la_k)\sinh(\la_j-\la_k+i\zeta)\sinh(\la_j-\la_k-i\zeta)}\nonumber\\
&\times
\vphantom{\int\limits_{-\infty}^{\infty}}
 \,\, Z_m(\{\la\}|\{\xi\})\,\,\det_m S(\{\la\},\{\xi\})\, H_s(\{\la\}|\{\xi\})\bar{H}_s(\{\la\}|\{\xi\}),
 \label{last_sum1}\end{align}
where  $\bar{H}_s$ means complex conjugation and the
 function $H_s(\{\la\}|\{\xi\})$ is defined as
\begin{equation}
H_s(\{\la\}|\{\xi\})=\frac{\pl_{j=1}^s\pl_{k=s+1}^m\sinh(\la_j-\la_k+i\zeta)
\pl_{1\le j<k\le s}\sinh(\la_j-\la_k)\pl_{s+1\le j<k\le m}\sinh(\la_j-\la_k)}
{\pl_{k=1}^m\left(\pl_{j=1}^s\sinh(\la_j-\xi_k-i\frac \zeta 2)\pl_{j=s+1}^m
\sinh(\la_j-\xi_k+i\frac \zeta 2)\right)}.
\end{equation}
It is easy to see that this function can be written as a Cauchy determinant
\begin{align*}
H_s(\{\la\}|\{\xi\})=&\pl_{j<k} \frac 1{\sinh(\xi_k-\xi_j)}\det \mathcal{H}^{(s)}, \\
 \mathcal{H}^{(s)}_{j k}= &\frac 1 {\sinh(\la_j-\xi_k-i\frac \zeta 2)}, \quad j\le s,\\
\mathcal{H}^{(s)}_{j k}= &\frac 1 {\sinh(\la_j-\xi_k+i\frac \zeta 2)}, \quad j > s.
\end{align*}
The sum over $s$ in  (\ref{last_sum1})  can be taken under the integrals. Here we separated the terms 
which depend on $s$:
\[\sul_{s=0}^m\frac {e^{\b s}}{s!(m-s)!}\det \mathcal{H}^{(s)}\det \bar{\mathcal{H}}^{(s)}.\]
This sum can be simplified if one introduce some auxiliary contour integrals:
\begin{align*}&\sul_{s=0}^m\frac {e^{\b s}}{s!(m-s)!}\det \mathcal{H}^{(s)}\det \bar{\mathcal{H}}^{(s)}\\
=\frac 1 {m!}
&\oint  \frac{d z_1}{2i\pi}\dots
\oint  \frac{d z_m}{2i\pi} 
\exp\left(\frac {\b}{i\zeta}\sul_{j=1}^m (z_j+i\frac \zeta 2)\right)\left(\pl_{j=1}^m 
\frac {\sinh 2z_j}{\sinh(z_j-i\frac \zeta 2)\sinh(z_j+i\frac \zeta 2)}\right)
\det \mathcal{F}^+\det \mathcal{F}^-,\end{align*}
where 
\[\mathcal{F}^\pm_{j k}= \frac 1{\sinh(\la_j\pm z_j-\xi_k)},\]
and contours are chosen  in such a way that the points $z_j=\pm i\frac \zeta 2$ are inside the contours
 and all the poles at the points $\la_j\pm z_j-\xi_k=0$ are outside. 
Thus we can represent the generating function as a single term
but the number of integrals is $2m$ now:
\begin{align}
\mathcal{Q}_m(\b)=& \frac 1{m!\sin^m\zeta}\pl_{j<k}
\frac 1{\sinh^3(\xi_j-\xi_k)}\int\limits_{-\infty}^{\infty}\! d\la_1\dots\!\!
\int\limits_{-\infty}^{\infty}\! d\la_m \oint \frac{d z_1}{2i\pi}\dots
\oint \frac{d z_m}{2i\pi} \,Z_m(\{\la\}|\{\xi\})\nonumber\\
&\times
\frac{\pl_{k=1}^m\pl_{j=1}^m\sinh(\la_j-\xi_k+i\frac \zeta 2)\sinh(\la_j-\xi_k-i\frac \zeta 2)}
{\pl_{ j> k}\sinh(\la_j-\la_k)\sinh(\la_j-\la_k+i\zeta)\sinh(\la_j-\la_k-i\zeta)}\, 
\,\det_m S(\{\la\},\{\xi\})\,\nonumber\\
&\times
\vphantom{\int\limits_{-\infty}^{\infty}}
 \,
\left(\pl_{j=1}^m 
\frac {\exp\left(\frac {\b}{i\zeta} (z_j+i\frac \zeta 2)\right)\sinh 2z_j}{\sinh(z_j-i\frac \zeta 2)\sinh(z_j+i\frac \zeta 2)}\right)
 \det \mathcal{F}^+\det \mathcal{F}^-.
\end{align}
Now, using  the symmetry of the expression under the 
integral with respect to permutations of pairs $(\la_j,z_j)$,
we can replace one of the determinants $\det \mathcal{F}^\pm$ by a product of its diagonal terms. 
It permits in particular to separate 
the variables $z_j$ and hence to integrate over them. 
It is easy to see that it leads to the following final result for the generating function:
\begin{align}
\mathcal{Q}_m(\b)=& \frac 1{\sin^m\zeta}\pl_{j<k}
\frac 1{\sinh^3(\xi_j-\xi_k)}\int\limits_{-\infty}^{\infty}\! d\la_1\dots\!\!
\int\limits_{-\infty}^{\infty}\! d\la_m \,Z_m(\{\la\}|\{\xi\})\,\det_m S(\{\la\},\{\xi\})\nonumber\\
&\times
\frac{\pl_{k=1}^m\pl_{j=1}^m\sinh(\la_j-\xi_k+i\frac \zeta 2)\sinh(\la_j-\xi_k-i\frac \zeta 2)}
{\pl_{ j> k}\sinh(\la_j-\la_k)\sinh(\la_j-\la_k+i\zeta)\sinh(\la_j-\la_k-i\zeta)}\, 
 \det_m \mathcal G ,\label{resum_final}
\end{align}
where the $m\times m$ matrix $\mathcal G$ is defined as 
\begin{equation}
\mathcal{G}_{j k}=\frac {e^\b} {\sinh(\la_j-\xi_k+i\frac \zeta 2)\sinh(\la_j-\xi_j-i\frac \zeta 2)}
+\frac 1{\sinh(\la_j-\xi_k-i\frac \zeta 2)\sinh(\la_j-\xi_j+i\frac \zeta 2)}.
\end{equation}
Similar results can be obtained for the two-point functions. This is the most compact formula for the
generating function (we reduced the number of terms from $2^m$ to one). However we think that the formulae
obtained in the 
 two previous sections are more convenient for the asymptotic analysis as they permit a natural 
homogeneous limit (which is not the case of (\ref{resum_final})).

\section{Free fermion point}

As the first application and check of this new re-summation formula we consider the free
 fermion point ($\zeta=\frac \pi 2$). Of course  the representations for 
the correlation functions in this point have already been 
obtained by different methods, but the formulae (\ref{resum_gff}) and (\ref{resum+-f}) 
give a simple and elegant way to get these explicit results.

\subsection{Generating function}

We calculate the generating function 
\[\mathcal{Q}_m(\b)\equiv \bra{\psi_g}\exp\{\b Q_{1,m}\}\ket{\psi_g}, 
\quad Q_{1,m}=\sul_{j=1}^m \frac 12(1-\s_j^z).\]

This function can be written as
\begin{equation}
\mathcal{Q}_m(\b)= \sul_{s=0}^m e^{s\b } F_s(m) ,
\label{sums}
\end{equation}
where contributions $F_s(m)$ are given by (\ref{resum_gff}).

Taking into account that $\zeta=\frac \pi 2$ all the determinants can be 
calculated. In the homogeneous limit one obtains the following multiple
integral representation:
\begin{align}
F_s(m)=&\frac {2^{m^2-m}}{\pi^m s!(m-s)!}
\int\limits_{-\infty}^{\infty}\! d\la_1\dots\!\!
\int\limits_{-\infty}^{\infty}\! d\la_m \pl_{1<j<k\le s}\sinh^2(\la_j-\la_k)
\nonumber\\
 &\times\pl_{s<j<k\le m}
\sinh^2(\la_j-\la_k)\pl_{j=1}^s\pl_{k=s+1}^m\cosh^2(\la_j-\la_k)
\pl_{j=1}^m\cosh^{-m}2\la_j.
\end{align}
After changing variables
\[\la_j=\frac 12 \log(\tan p_j),\]
we obtain a much simpler representation:
\begin{equation}
F_s(m)=\frac {2^{m^2-m}}{\pi^m s!(m-s)!}
\int\limits_{0}^{\frac \pi 2} d p_1\dots
\int\limits_{0}^{\frac \pi 2} d p_m\pl_{j>k}\sin^2(\varepsilon_j p_j-
\varepsilon_k p_k).
\end{equation}
The expression under the integral can be rewritten as a product of two Vandermonde
determinants:
 \begin{align*}
\pl_{j>k}\sin^2(\varepsilon_j p_j-
\varepsilon_k p_k)=&2^{m-m^2}
\left|\pl_{j>k}(e^{2i\varepsilon_j p_j}-e^{2i\varepsilon_k p_k})\right|^2
=2^{m-m^2}\det V(\{\varepsilon p\})\det V^*(\{\varepsilon p\}),\\
V_{j k}(\{p\})=&e^{2i (k-1) p_j},
\end{align*}
where star means hermitian conjugation.The product of these two determinants
can be calculated as the determinant of the product of two matrices:
\begin{align*}
\det V(\{\varepsilon p\})\det V^*(\{\varepsilon p\})&=
\det (V^*(\{\varepsilon p\})V(\{\varepsilon p\}))\\
(V^* V)_{kn}=&\sul_{j=1}^m e^{2i (k-n)\varepsilon_j p_j}.
\end{align*}
Now we should take into account that we integrate this determinant. It can be 
written as a sum over permutations,
\begin{equation*}
F_s(m)=\frac {1}{\pi^m s!(m-s)!}
\int\limits_{0}^{\frac \pi 2} d p_1\dots
\int\limits_{0}^{\frac \pi 2} d p_m\sum_{\sigma}\det W(\sigma,s,\{p\}), 
\end{equation*}
where the matrix $W$ is defined as 
\[ W_{k n}(\sigma,s,\{p\})=e^{2i (k-n)\varepsilon_{\sigma(n)} p_{\sigma(n)}}. \]
As we integrate over all the variables $p$, permutations of the variables
with the same value of the  parameter $\varepsilon$ do not change the 
integration result and thus the sum over permutations can be replaced by a sum
over the partitions of the set $\{p\}$ into two subsets  $\{p^+\}$ and 
$\{p^-\}$, with a number of elements in the first one being $s$,
  \begin{equation*}
F_s(m)=\frac {1}{\pi^m}
\int\limits_{0}^{\frac \pi 2} d p_1\dots
\int\limits_{0}^{\frac \pi 2} d p_m\sum_{\{p\}=\{p^+\}\cup\{p^-\}}\det 
\tilde{W}(\{p^+\},\{p^-\}). 
\end{equation*}
  Here the matrix $\tilde{W}(\{p^+\},\{p^-\})$ is defined as 
\[\tilde{W}_{k n}(\{p^+\},\{p^-\})=e^{2i (k-n)\epsilon_{n} 
p^{\epsilon_n}_{n}},\]
$\epsilon_n$ being $+$ or $-$.

 Now considering the entire sum over $s$ (\ref{sums}),
 one can easily note that it can be rewritten as a determinant of a sum 
of two matrices:
\begin{align}
\mathcal{Q}_m(\b)=&\frac {1}{\pi^m}
\int\limits_{0}^{\frac \pi 2} d p_1\dots
\int\limits_{0}^{\frac \pi 2} d p_m\det_m U(\b,\{p\}),\nonumber\\
U_{k n}(\b,\{p\})=&e^{2i (k-n)p_n}+e^\b e^{2i (n-k)p_n}.
\end{align}
It is now possible to calculate all the integrals and to write the final result as 
a determinant:
\begin{align}
\mathcal{Q}_m(\b)=&\det_m T(\b),\nonumber\\
T_{k n}(\b)=&\delta_{k n}\frac{e^\b+1}2+
(1-\delta_{k n})(1-e^\b)\frac {1-(-1)^{n-k}}{2i\pi(n-k)}.
\label{res_gf_ff}
\end{align}
 From this formula one can easily obtain the well known result for the two-point function
$g_{z z} (m)$:
\begin{equation}
\langle\sigma_1^z\sigma_{m+1}^z\rangle=
\frac2{\pi^2m^2}\Bigl((-1)^m-1\Bigr).
\end{equation}
 The emptiness formation probability \cite{ShiTN01} 
comes directly from (\ref{res_gf_ff}) by taking the limit 
$\b\rightarrow -\infty$.

\subsection{Two-point functions}

The two-point functions
$\l\sigma_1^+\sigma_{m+1}^-\r$ and $\l\sigma_1^-\sigma_{m+1}^+\r$
can be calculated in a similar way in the free fermion point.

First of all a re-summation formula can be obtained for these functions
almost in the same way as for the generating function. We can consider
even a more general function:
\[g_{+-}(m,\b)=\l  \sigma_1^+\exp (\b Q_{2,m})\sigma_{m+1}^-\r,\]
where $Q_{2,m}=\sul_{j=2}^m \frac 12(1-\s_j^z)$,
which includes also the correlation function of fermionic fields.
After re-summation it can be represented as:
\begin{equation}
g_{+-}(m,\b)= \sul_{s=0}^{m-1} e^{s\b } g^{+-}_s(m) ,
\label{sums+-}
\end{equation}
where the contributions $g^{+-}_s(m)$ are given by (\ref{resum+-f}).
This  general re-summation formula
 can be  simplified in the free-fermion case:
\begin{align}
g_s^{+-}(m)=&\frac {2^{m^2+1}(-1)^s}{i^{m-1}\pi^{m+1} s!(m-1-s)!}
\int\limits_{-\infty}^{\infty}\! d\la_1\dots\!\!
\int\limits_{-\infty}^{\infty}\! d\la_m 
\int\limits_{-\infty}^{\infty}\! d\la_+
\int\limits_{-\infty}^{\infty}\! d\la_-
\nonumber\\
 &\times
 \frac{ \pl_{2<j<k\le s+1}\sinh^2(\la_j-\la_k)
\pl_{s+1<j<k\le m}
\sinh^2(\la_j-\la_k)\pl_{j=1}^{s+1}\pl_{k=s+2}^m\cosh^2(\la_j-\la_k)}
{\pl_{j=2}^{m}\cosh^{m}2\la_j}\nonumber\\
&\times\frac{\cosh(\la_+-\la_-)\!\pl_{j=2}^{s+1}\!\sinh(\la_+-\la_j)
\cosh(\la_--\la_j)\!\pl_{j=s+2}^{m}\!\cosh(\la_+-\la_j)
\sinh(\la_--\la_j)}{\cosh^{m+1}2\la_+\cosh^{m+1}2\la_-}\nonumber\\
&\vphantom{\frac{\pl_1^m}{\pl_1^m}}
\times\sinh^{m}(\la_+-i\frac\pi 4)\sinh^{m}(\la_-+i\frac\pi 4).
\end{align}
Changing variables as in the previous case \[\la_j=\frac 12 \log(\tan p_j),\]
we obtain:
\begin{align}
g_s^{+-}(m)=&\frac {2^{m^2-m+1}(-1)^s}{i^{m-1}\pi^{m+1} s!(m-1-s)!}
\int\limits_{0}^{\frac \pi 2} d p_2\dots
\int\limits_{0}^{\frac \pi 2} d p_m
\int\limits_{0}^{\frac \pi 2} d p_+
\int\limits_{0}^{\frac \pi 2} d p_- e^{i(p_+-p_-)m}\nonumber\\
&\times
\pl_{m\ge j>k>1}\sin^2(\varepsilon_j p_j-
\varepsilon_k p_k)\pl_{j=2}^m\sin(p_+-\varepsilon_j p_j)
\sin(p_-+\varepsilon_j p_j)\sin(p_++p_-).
\end{align}
This expression also can be rewritten as a product of two Vandermonde 
determinants, but now the matrices have different sizes:
 \begin{align*}
g_s^{+-}(m)=&\frac {(-1)^s}{i^{m}\pi^{m+1} s!(m-1-s)!}
\int\limits_{0}^{\frac \pi 2} d p_2\dots
\int\limits_{0}^{\frac \pi 2} d p_m
\int\limits_{0}^{\frac \pi 2} d p_+
\int\limits_{0}^{\frac \pi 2} d p_-e^{-2i\sum_{j=2}^m\varepsilon_j p_j}\\
&\times
\det_{m+1} V(m+1,\{p_+,-p_-,\varepsilon_2 p_2,\dots, \varepsilon_m p_m\})
\det_{m-1} V^*(m-1\{\varepsilon_2 p_2,\dots, \varepsilon_m p_m\}),\\
V_{j k}=\vphantom{\frac{\pl_{t}^m}{\pl_{j}^j}}&e^{2i (k-1)\varepsilon_j p_j}.
\end{align*}
The product of these two determinants can be again rewritten as a determinant
of a product of two matrices if we add two rows and two columns to the 
second one:
\begin{align*}
\tilde{V}^*_{1j}(m+1,\{p\})=&\tilde{V}^*_{j1}(m+1,\{p\})=\delta_{1j},\\
\tilde{V}^*_{2j}(m+1,\{p\})=&\tilde{V}^*_{j2}(m+1,\{p\})=\delta_{2j},\\
\tilde{V}^*_{k j}(m+1,\{p\})=&V^*_{k-2\,j-2 }(m-1,\{p\}),\quad k,j>2.
\end{align*}
Now we can take the product of these two determinants and proceed with the sum over
$s$ in the same way as for the generating function: 
\begin{align}
g_{+-}(m,\b)=&\frac {1}{i^{m}\pi^{m+1}}
\int\limits_{0}^{\frac \pi 2} d p_2\dots
\int\limits_{0}^{\frac \pi 2} d p_m\int\limits_{0}^{\frac \pi 2} d p_+
\int\limits_{0}^{\frac \pi 2} d p_-
\det_{m+1} \tilde{U}(\b,\{p\}),\nonumber\\
\tilde{U}_{1 n}(\b,\{p\})=&e^{2i (n-1)p_+},\nonumber \\
\tilde{U}_{2 n}(\b,\{p\})=&e^{-2i (n-1)p_-}, \nonumber\\
U_{k n}(\b,\{p\})=&e^{2i (k-n-1)p_n}-e^\b e^{2i (n-k+1)p_n}.
\end{align}
Now all the integrals can be calculated. For the two-point function ($\b=0$)
we get
\begin{align}
g_{+-}(m)=& \frac {(-1)^m} {\pi^{m+1}} \det_{m+1}T^{-+}(m),\nonumber\\
T^{-+}_{1 1}=&T^{-+}_{2 1}= \frac\pi 2,\nonumber\\
T^{-+}_{1 n}=&-T^{-+}_{2 n}=\frac{1+(-1)^n}{2(n-1)},\nonumber\\
T^{-+}_{k n}=&\frac{1+(-1)^{k-n}} {(k-n-1)}.
\end{align}
This determinant can be easily computed and gives a formula obtained  by Wu in \cite{Wu66} (a more general formula was obtained by McCoy \cite{Mcc68}):
\be{LASS}
\langle\sigma_1^+\sigma_{m+1}^-\rangle=\frac{(-1)^m}2
\prod_{k=1}^{\left[\frac{m}2\right]}
\frac{\Gamma^2(k)}{\Gamma(k-\frac12)\Gamma(k+\frac12)}
\prod_{k=1}^{\left[\frac{m+1}2\right]}
\frac{\Gamma^2(k)}{\Gamma(k-\frac12)\Gamma(k+\frac12)}.
\ee
The asymptotic behaviour of such a product was also obtained in \cite{Wu66} using the technique introduced in \cite{Bar}:
\be{LAresas}
\langle\sigma_1^+\sigma_{m+1}^-\rangle=
\frac{(-1)^m}{\sqrt{2m}}\exp\left\{
\frac12\int_0^\infty\frac{dt}t\left[e^{-4t}-
\frac1{\cosh^2t}\right]\right\}
\left(1-\frac{(-1)^m}{8m^2}+{\cal O}(m^{-4})\right).
\ee

{\bf Acknowledgements}\\
J. M. M., N. S. and V. T. are supported by CNRS. N. K., J. M. M.,
V. T. are supported by the European network
EUCLID-HPRNC-CT-2002-00325.  J. M. M. and N.S. are supported
by INTAS-03-51-3350. N.S. is supported
by the French-Russian Exchange Program, the Program of RAS
Mathematical Methods of the Nonlinear Dynamics,
RFBR-02-01-00484, Scientific Schools 2052.2003.1.
N. K, N. S. and V. T. would like to thank the Theoretical Physics group
of the Laboratory of Physics at ENS Lyon for hospitality, which makes
this collaboration possible. We are also grateful to the orginizers
of the RIMS COE 2004 research programm in Kyoto where this article was completed.

\bibliographystyle{h-elsevier} 

%

\end{document}